\documentclass[aps,floatfix,twocolumn,showpacs]{revtex4}
\usepackage{natbib}
\usepackage{graphicx}
\usepackage{graphics,epsfig}
\usepackage{amssymb}
\newcommand{\physrep}{Phys.~Rep.}
\newcommand{\araa}{Ann.~Rev.~Astron.~Astrophys.}

\begin{document}
\title{Probing Cosmic Acceleration Beyond the Equation of State: Distinguishing between Dark Energy and Modified Gravity Models}
\author{Mustapha Ishak$^{1,2}$\cite{email},}
\author{Amol Upadhye$^{3}$,}
\author{David N. Spergel$^{2}$}
\affiliation{
$^1$ Department of Physics, The University of Texas at Dallas, Richardson, TX 75083, USA\\
$^2$ Department of Astrophysical Sciences, Princeton University,
  Princeton, NJ 08544, USA \\
$^3$ Department of Physics, Princeton University,
  Princeton, NJ 08544, USA }
\date{\today}
\begin{abstract}
If general relativity is the correct theory of physics on large scales,
then there is a differential equation that relates the Hubble expansion 
function, inferred from measurements of angular diameter distance 
and luminosity distance, to the growth rate of large scale structure. 
For a dark energy fluid without couplings 
or an unusual sound speed, deviations 
from this consistency relationship could be the signature of modified 
gravity on cosmological scales. We propose a procedure based on this 
consistency relation in order to distinguish between some dark energy models and modified 
gravity models. The procedure uses different combinations of cosmological 
observations and is able to find inconsistencies when present. 
 As an example, we apply the procedure to a 
universe described by a recently proposed 5-dimensional modified gravity model. 
We show that this leads to an inconsistency within the dark
energy parameter space detectable by future experiments. 
\end{abstract}
\maketitle
%
%%%%%%%%%%%%%%%%%%%%%%%%%%%%%%%%%%%%%%%%%%%%%%%%%%%%%%
 
\section{Introduction.}
\label{sec:introduction}
Several cosmological observations, e.g. 
\cite{observations1,observations2,observations3,observations4,observations5,observations6,observations7,observations8}, 
have demonstrated that the expansion of the universe has entered a phase of
acceleration. The cosmic acceleration and the questions associated
with it constitute one of the most important and challenging problems
 for several fields of physics (astrophysics, gravitation, high energy 
physics, and fundamental physics), e.g. 
\cite{reviews1,reviews2,reviews3,reviews4,reviews5,reviews6}. 
A dark energy component that
represents 2/3 of the entire energy content of the universe has
been proposed to explain cosmic acceleration 
\cite{quintessence1,quintessence2,quintessence3,quintessence4}. 
This dark energy component can lead to repulsive gravity because of 
its negative equation of state. 

Besides dark energy, several models 
based on modifications to the gravity sector have been proposed 
recently as alternatives to explain the cosmic acceleration, 
e.g. \cite{DGP,modifiedgravity1,modifiedgravity2}. 

These two families of models, dark energy and modified gravity,  
are fundamentally different, and  a question 
of major importance is to distinguish between the two possibilities
 using cosmological data. 
In this paper, we propose a procedure that uses different pairs of 
cosmological observations in order to address this question. The procedure is 
able to find inconsistencies in the dark energy parameter space 
due to an underlying modified gravity model. 

Furthermore, a burning question that comes to mind after an equation of state 
of dark energy has been determined from cosmological observations is as follows: 
Is this really a dark energy equation of state or just a results 
obtained because one tried to fit a dark energy model on the top of some 
modified gravity model? The procedure proposed also addresses this question.

The outline of the paper is as follows. In section two, we introDuce the basic 
idea of the consistency test. Next, in section three, we recall some of the 
commonly used parameterizations of dark energy and also an example of modified 
gravity models. In section four, we discuss constraints on the expansion history 
while in section five we discuss constraints on the growth rate of large scale 
structure. Section six describes our approach. In section seven, we describe our 
implementation of the procedure using cosmological observations and present our results. 
We discuss our results and conclude in the last section.
%
%%%%%%%%%%%%%%%%%%%%%%%%%%%%%%%%%%%%%%%%%%%
 
\section{General Relativity consistency relationship.
  } 
Einstein's equations of General Relativity relate the curvature of 
spacetime to the matter and energy content of the universe, thus describing 
the cosmological dynamics. 
Cosmic acceleration affects both the expansion history of the
universe (given by the Hubble function $H(z) \equiv \left. \frac{1}{a} \frac{da}{dt}
\right|_{a=(z+1)^{-1}}$, where \textit{a} is the scale factor and \textit{z} is the redshift) 
and the rate at which clusters of galaxies grow (given by the growth rate 
of large scale structure, $D=\delta(a)/\delta(a_i)$, where $\delta(a)\equiv 
\frac{\rho-\bar{\rho}}{\bar{\rho}}$ is the overdensity). 

Importantly, the 
expansion function must be consistent with the growth rate function via 
Einstein's equations. If the cosmic acceleration is not due to a dark energy 
component in Einstein's equations then the presence of significant 
deviations from the consistency relation between the expansion and the 
growth rate will be a symptom of a breakdown of General Relativity at 
cosmological scales and a hint to possible modified gravity models at very large scales.

In this paper, the consistency test is implemented using 
specific combinations of simulated future cosmological observations.

%
%%%%%%%%%%%%%%%%%%%%%%%%%%%%%%%%%%%%%%%%%%%
 
\section{Dark energy models versus 
  Modified Gravity.} 
\label{sec:acceleration_models}
\subsection{Dark Energy parameterization} We assume that dark energy can be 
described in terms of a cosmic fluid, with an equation of state 
$w = P / \rho$ that can vary with redshift.  
We parameterize the variation of $w(z)$ using: 
\begin{itemize}
\item{$w(z) = w_0 + w_1 z$ if $z<1$ and $w(z)= w_0 + w_1$ otherwise, e.g.  
\cite{observations4,Upadhye}, and also the Taylor expansion}
\item{$w(a)= w_0 + w_a (1-a) + w_b (1-a)^2$, e.g. \cite{Chevallier,Linder}.} 
\end{itemize}
Given $w(z)$, the dark energy density as a function of redshift
can be described using the dimensionless function 
\begin{equation}
{\mathcal{Q}}(z) \equiv \rho_{de}(z) / \rho_{de}(0)=\exp{3\int_{0}^{z}\frac{1+w(z'))dz'}{1+z'}}.
\end{equation} 
We assume a spatially flat universe and do not include massive neutrinos 
or isocurvature perturbations.
For supernova (SN Ia) calculations, we use the parameters
\begin{equation}
p^\alpha=\{\Omega_\Lambda,w_0,w_1,\mathcal{M} \}.
\end{equation} 
For weak gravitational 
lensing (WL) calculations, we use: 
\begin{equation}
p^\alpha=\{\Omega_m h^2,\Omega_\Lambda,w_0,w_1,n_s,\sigma_8^{lin},z_p,\xi_s,\xi_r\}.
\end{equation} 
For CMB, we use: 
\begin{equation}
p^\alpha=\{\Omega_m h^2,\Omega_b h^2,
\Omega_\Lambda,w_0,w_1,n_s,\sigma_8^{lin},\tau\}
\end{equation} 
where
$\Omega_{de} \equiv \rho_{de}(0) / \rho_{crit}(0)$ is the dark energy
density fraction, and $\rho_{crit}(0) \equiv {3 H_0^2}/{8 \pi
  G}$; 
 $\Omega_m h^2$ is the physical matter density;
 $\Omega_b h^2$ is the physical baryon density; 
 $\tau$ is the optical depth to the surface of last scattering; 
 $\sigma_8^{lin}$ is the amplitude of linear fluctuations; 
 $n_s$ is the spectral index of the primordial power spectrum;
 $z_p$ is the characteristic redshift of source galaxies for weak
lensing \cite{Ishaketal2004,IshakandHirata2005};
 $\xi_s$ and $\xi_r$ are the lensing absolute and relative calibration parameters 
as defined in \cite{Ishaketal2004}.

\subsection{Modified gravity models}
In this paper, we are interested in using a good example of modified gravity 
model but only in order to illustrate the procedure proposed. We are not interested 
here in evaluating any particular model of modified gravity.

Although several models of modified gravity have been proposed 
recently as alternatives to dark energy, we chose in this analysis to use the 
Dvali-Gabadadze-Porrati (DGP) model. The model is inspired 
by higher dimensional physics and is  consistent with 
current observations, e.g. \cite{Deffayet3,Sawicki}. As mentioned, 
our interest is to use the DGP model as an example of deviation 
from General Relativity in order to test the proposed procedure.
We refere the reader to studied dedicated to the DGP model
phenomenology \cite{Deffayet3,Sawicki,LueReport,Fairbairn}

We provide in this sub-section a brief
introduction to the model, but we refer the reader to \cite{DGP,
  DeffayetDvaliGabadadze2002} for a full description. The action for
the five-dimensional theory is \cite{DGP, DeffayetDvaliGabadadze2002}
\begin{eqnarray}
S_{(5)} &=& \frac{1}{2}M_{(5)}^3 \int d^4x \, dy \sqrt{-g_{(5)}}
R_{(5)} \nonumber \\  &+&  \frac{1}{2}M_{(4)}^2 \int d^4x \sqrt{-g_{(4)}} R_{(4)}
+ S_{matter}, 
\end{eqnarray}
where the subscripts $4$ and $5$ denote quantities on
the brane and in the bulk, respectively; $M_{(5)}$ is the five
dimensional reduced Planck mass; $M_{(4)}=2.4 \times 10^{18}
\textrm{GeV}$ is the four dimensional effective reduced Planck mass;
$R$ and $g$ are the Ricci scalar and the determinant of the metric,
respectively.  The first and second terms on the right hand side
describe the bulk and the brane, respectively, while $S_{matter}$ is
the action for matter confined to the brane. 
The two different prefactors $M_{(5)}^3/2$ and $M_{(4)}^2/2$ in front
of the bulk and brane actions give rise to a characteristic length,
scale \cite{Deffayet2001} 
\begin{equation}
r_c = {M_{(4)}^2}/{2 M_{(5)}^3}.
\end{equation}
 If $M_{(5)}$ is much less than $M_{(4)}$, then the brane terms in the
action above will dominate over the bulk terms on
scales much smaller than $r_c$, and gravity will appear four
dimensional.  For example, nonrelativistic small-scale gravity will
obey the Newtonian inverse-square force law.  On scales larger than
$r_c$, the full five dimensional physics will be recovered, and the
gravitational force law will revert to its five dimensional $1/r^3$
form.  This is usually discussed in terms of gravity leakage into an extra 
dimension. Ref. \cite{DeffayetLandauRaux} shows that tuning $M_{(5)}$  to
about $10-100 \textrm{MeV}$, implying $r_c \sim H_0^{-1}$, is
consistent with cosmological data. They have also been discussed 
in \cite{lue}. 
Low redshift cosmology in DGP brane worlds was studied in
\cite{Deffayet2001,DeffayetDvaliGabadadze2002}.  Defining the
effective energy density 
\begin{equation}
\rho_{r_c} \equiv \frac{3}{(32 \pi G r_c^2)},
\end{equation}
Friedmann's first equation becomes 
\begin{equation}
H_{_{DGP}}^2 + \frac{k}{a^2} = \frac{8 \pi
  G}{3} \left(\sqrt{\rho + \rho_{r_c}} + \sqrt{\rho_{r_c}}
\right)^2.
\end{equation}
 For each perfect fluid $i$ on the brane, we may define
$\Omega_i \equiv \rho_{i,0} / \rho_{crit,0}$.
Following \cite{DeffayetDvaliGabadadze2002}, we also define
$\Omega_{r_c} \equiv \frac{1}{4} r_c^{-2} H_0^{-2}$. 
We focus here on a flat universe containing only nonrelativistic
matter, such as baryons and cold dark matter, in which 
$\Omega_{r_c} = \left( \frac{1-\Omega_m}{2} \right)^2$.
In this model, the gravitational ``leakage'' into the fifth 
dimension, on large length scales, becomes a substitute for 
dark energy.  
%
%%%%%%%%%%%%%%%%%%%%%%%%%%%%%%%%%%%%%%%%%%%%%%%%%%
 
\section{Constraints from the expansion history using supernovae of type Ia.}
{
\begin{figure}[t]
\begin{center}
\includegraphics[width=2.5in,height=3.2in,angle=-90]{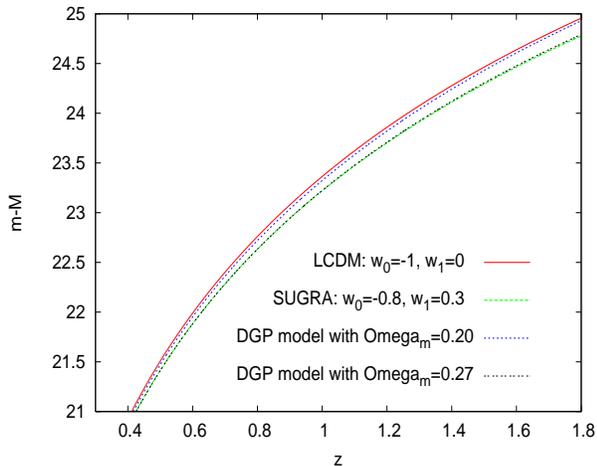}
\caption{\label{fig:hubbles} 
Supernova Hubble diagrams for several dark energy and DGP models. 
 Note that the $\Lambda$CDM model (red solid line) and the $\Omega_m=0.20$ DGP 
model (blue dotted) have nearly identical Hubble diagrams, but different 
growth factors as shown in Fig.\ref{fig:growth1}.  The same is true of the 
SUGRA (green dashed) and $\Omega_m=0.27$  DGP (black double dotted) models. }
\end{center}
\end{figure}
}
For a spatially flat universe with dark energy, the 
Hubble parameter $H(z)$ can be expressed in terms of its current 
value $H_0$ and the function ${\mathcal{Q}}(z)$ as 
\begin{equation}
H_{_{DE}}(z) = H_0
\sqrt{\Omega_m (1+z)^3 + (1-\Omega_m) {\mathcal{Q}}(z)}.  
\end{equation}
Since probes such as the SN Ia surveys give information about the redshift 
variation of $H(z)$ but not its current value, we define the
dimensionless Hubble parameter for dark energy models as  
\begin{equation}
{\mathcal{H_{DE}}}(z) \equiv H(z)/H_0 = \sqrt{\Omega_m (1+z)^3 +
  (1-\Omega_m) {\mathcal{Q}}(z)}. 
\label{eqn:calHde}
\end{equation}
{
\begin{figure}[t]
\begin{center}
\includegraphics[width=2.5in,height=3.2in,angle=-90]{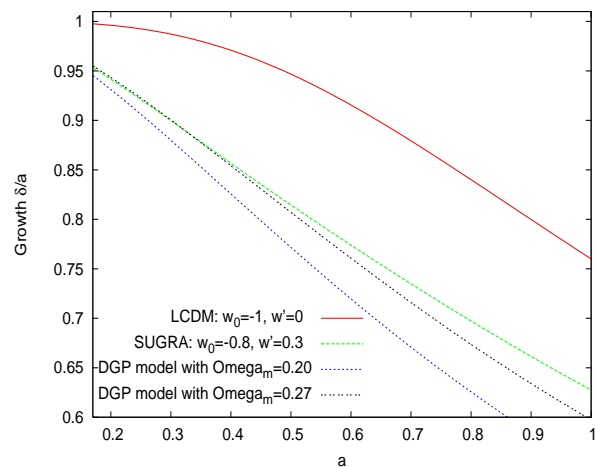}
\caption{\label{fig:growth1} 
Growth factors of linear density perturbations for several dark energy and DGP models. 
Comparisons among several dark energy and DGP models growth factors of linear density perturbations. Note that the growth factor in the $\Omega_m=0.27$ DGP model is suppressed with respect to that in the $\Lambda$CDM model, which has the same $\Omega_m$. }
\end{center}
\end{figure}
}
For spatially flat DGP model, the expansion is given
by  
\begin{equation}
\label{eqn:HHdgp}
{\mathcal{H_{DGP}}}(z) = \frac{1}{2}(1-\Omega_m) +
\sqrt{\frac{1}{4}(1-\Omega_m)^2 + \Omega_m (1+z)^3}. 
\end{equation}
To probe the expansion history, we consider calibrated type Ia
supernovae as standard candles to measure the luminosity distance as
a function of the redshift. Recall that the SN Ia apparent
magnitude as a function of redshift is given by 
\begin{equation}
m(z) = 5 \log_{10} \left( D_L(z) \right) + {\mathcal{M}}.
\end{equation}
Here, ${\mathcal{M}}$ depends on $H_0$ as well as the absolute
magnitude of type Ia supernovae.  We treat ${\mathcal{M}}$ as a
nuisance parameter.  Meanwhile, the dimensionless luminosity distance
is 
\begin{equation}
D_L(z)=(1+z) \int_0^z dz' / {\mathcal{H}}(z').
\end{equation}
As well known, $m(z)$ depends on $w_0$ and $w_1$ only through a
complicated integral relation, which ``smears out'' dark energy
information \cite{maor2001}.  The result is a degeneracy which allows very different
models to have nearly identical $m(z)$.  
Fig.\ref{fig:hubbles}
shows $m(z)$ for a few dark energy and DGP cosmologies. 
The figure shows that some DGP models are very degenerate with certain 
dark energy models, such as $\Lambda$CDM models or SUGRA-inspired models. 
Evident in the expression for $m(z)$ is another degeneracy, between $\Omega_m$ 
and the equation of state parameters.  We break this degeneracy by combining SN Ia 
data with the CMB power spectrum.
We assume the availability of a sample of 2000 
supernovae, evenly distributed in redshift between
$z_{Min}=0$ and $z_{Max}=1.7$, with a magnitude uncertainty per
supernova of $\sigma_m=0.2$.  As in \cite{Upadhye,Ishak2005}, we also
include a peculiar velocity uncertainty $\sigma_v=500$km/s, and a
systematic uncertainty $\delta m = 0.02$ in bins of 
$\Delta z = 0.1$. 
%
%%%%%%%%%%%%%%%%%%%%%%%%%%%%%%%%%%%%%%%%%%%%%%%%%%
 
\section{Constraints from the growth of large scale
  structure and gravitational lensing tomography.} 
{
\begin{figure}[t]
\begin{center}
\includegraphics[width=2.5in,height=3.2in,angle=-90]{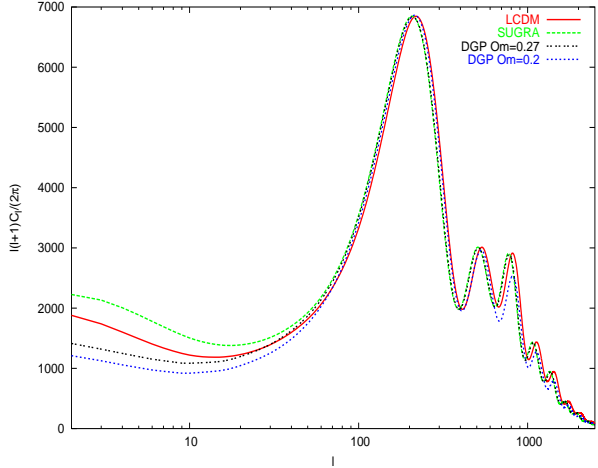}
\caption{\label{fig:cmbCl} 
CMB power spectra for several dark energy and DGP models: 
$\Lambda$CDM model is in red solid line;
SUGRA model is in green dashed line;
$\Omega_m=0.27$ DGP model is in black double dotted line; 
$\Omega_m=0.20$ DGP model is in blue dotted line;
Following \cite{Deffayet3}, a modified version of {\sc CMBFAST} \cite{Zaldarriaga} was used to generate the DGP plots.
}
\end{center}
\end{figure}
}
For dark energy models, the suppression of the growth of large scale
structure is due to an increase in the ratio of dark energy density
to matter density. We calculate the growth factor for 
these models by numerical integration of the differential equation from 
\cite{Ma1999,linder2003},
\begin{equation}
G''+\left[\frac{7}{2}-\frac{3}{2}\frac{w(a)}{1+X(a)}\right]\frac{G'}{a}+\frac{3}{2}\frac{1-w(a)}{1+X(a)}\frac{G}{a^2}=0,  
\label{eqn:growthDE}
\end{equation}
where $'\equiv d/da$, $G=D/a$ is the normalized growth factor,  
\begin{equation}
X(a)=\frac{\Omega_m}{(1-\Omega_m)a^3 \mathcal{Q}(a)}, 
\end{equation}
and $\mathcal{Q}(a)$ is as given earlier.

For DGP models, the suppression of the growth is due to weakened
gravity resulting from ``leakage'' into the extra dimension. Following
\cite{lue,Koyama}, we use
\begin{equation}
\ddot{\delta}+2 H_{_{DGP}} \dot{\delta}- 4 \pi G \rho \big{(}1+\frac{1}{3\beta}\big{)} \delta =0
\label{eqn:growthDGP}
\end{equation}
where 
\begin{equation}
\beta=1-2r_c H_{_{DGP}} \Big{(}1+\frac{\dot{H}_{_{DGP}}}{3 H_{_{DGP}}^2}\Big{)}
\end{equation}
Fig.\ref{fig:growth1} shows that, compared to a $\Lambda$CDM model with
the same matter density, a DGP model has a distinct suppression of the
growth factor. Also, Fig.\ref{fig:growth1} displays how the
degenerate models of Fig.\ref{fig:hubbles} show distinct growth
factor functions. In the next two sections, we will take advantage of
this difference in order to detect possible signatures of DGP models.

{
\begin{figure}[t]
\begin{center}
\includegraphics[width=2.5in,height=3.2in,angle=-90]{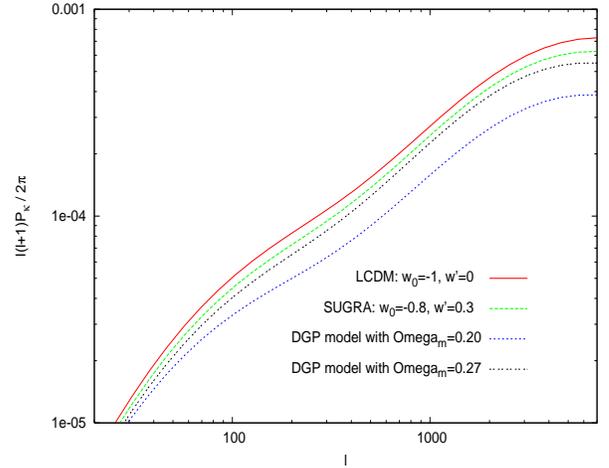}
\caption{\label{fig:convergences} 
Lensing convergence power spectra for several dark energy and DGP models: 
$\Lambda$CDM model is in red solid line;
SUGRA model is in green dashed line;
$\Omega_m=0.27$ DGP model is in black double dotted line; 
$\Omega_m=0.20$ DGP model is in blue dotted line;
}
\end{center}
\end{figure}
}

We consider weak gravitational lensing (cosmic shear) 
tomography as a probe of the growth factor of large scale structure. 
Indeed, cosmic shear observations capture the effect of the cosmic 
acceleration on both the expansion and the growth factor. 
We follow the formalism and conventions as used in 
\cite{Ishaketal2004,Ishak2005} and we use the convergence power
spectrum as our statistic
\cite{1992ApJ...388..272K,1997ApJ...484..560J,1998ApJ...498...26K}, 
\begin{equation}
P^{\kappa}_l =
\frac{9}{4} H_0^4\Omega_m^2\int^{\chi_H}_{0}
\frac{g^2(\chi)}{a^2(\chi)}P_{3D}
\left(\frac{l}{\sin_{K}(\chi)},\chi\right) d\chi,
\end{equation}
where $P_{3D}$ is the $3D$ nonlinear power spectrum of the matter
density fluctuation, $\delta$;  $a(\chi)$ is the scale factor; and
$\sin_{K}(\chi)=K^{-1/2}\sin(K^{1/2}\chi)$ is the comoving angular
diameter distance to $\chi$ (for the spatially flat universe used in
this analysis, this reduces to $\chi$). The weighting function
$g(\chi)$ is the source-averaged distance ratio given by 
\begin{equation}
g(\chi) =
\int_\chi^{\chi_H} n(\chi') {\sin_K(\chi'-\chi)\over \sin_K(\chi')}
d\chi'
\end{equation}
 where $n(\chi(z))$ is the source redshift distribution
(we use the distribution
\cite{2000Natur.405..143W} $n(z)=\frac{z^2}{2 z_0^3}\, e^{-z/z_0}$,
which peaks at $z_p=2z_0$). We integrate
numerically the growth factor function given by
Eqns. \ref{eqn:growthDE} or \ref{eqn:growthDGP} above, and we use the
linear-to-nonlinear mapping procedure {\sc halofit} \cite{smith2003}.
We used the {\sc halofit} for the DGP models as well because we are only
interested into using them as an example of deviation from general relativity
in order to test our procedure. Dedicated studies to DGP phenomenology
should consider using a better approximation for these models. 
As mentioned above, the expansion history is contained in the window
function, while the growth factor is contained in the $3D$ nonlinear
matter power spectrum.  
Fig.\ref{fig:convergences} shows convergence power spectra
corresponding to dark energy models and modified gravity (DGP) models.

The separation of source galaxies into tomographic bins improves
significantly the constraints on cosmological parameters, and
particularly those of dark energy \cite{Hu1,Hu2}.  
The constraints obtained from different bins
are complementary, and add up to reduce the final
uncertainties on the parameters. Based on recently proposed 
surveys (e.g. \cite{Albert2005}), 
we use a reference survey with a sky coverage of 10\%,
a source density $\bar n =100\;{\rm gal/arcmin}^2$, an rms 
intrinsic ellipticity $\left<\gamma_{int}^2\right>^{1/2} \approx 0.25$, and a
median redshift $z_{med}=1.5$.
We assume a reasonable photometric redshift uncertainty 
of $\sigma(z)=0.05$, and we split the sources into 10 tomographic 
bins over the range $0.0 < z < 3.0$ with intervals of
$\Delta z=0.3$. 
We have used $\ell_{\rm max}=3000$, since the assumption
of a Gaussian shear field, underlying the Fisher formalism 
and the {\sc halofit} approximation,   
 may not be valid for larger $\ell$s. For the minimum
$\ell$, we take the fundamental mode approximation, i.e., we consider
only lensing modes for which at least one wavelength can fit inside
the survey area.  
%
%%%%%%%%%%%%%%%%%%%%%%%%%%%%%%%%%%%%%%%%%%%%%%%%%%
 
\section{The fundamentals of the consistency test: 
Constraints from the expansion versus constraints from the growth of LSS}
The cosmic acceleration affects cosmology in two ways:
i) It affects the expansion history of the universe and 
ii) It supresses the growth rate of large scale structure in the universe.  
 The idea that we explore is that, for dark energy models, these two effects must be consistent one with another because they are related by General Relativity. 
The presence of significant inconsistencies between the expansion history and the growth rate would be the signature of some modified gravity models at cosmological scales. We propose a procedure that detects such inconsistencies when they are present. 

The approach we explore is as follows.
We assume that the true cosmology is described by a DGP model, and 
then ask what contradictions arise when the data are instead analyzed 
based on the assumption of a dark energy model (as mentioned earlier, we are not
particularly interested in the DGP cosmology here but we want to use it as 
an example to illustrate the procedure.) 

Because we will generate the 
data using the DGP model, the consistency relation from General Relativity
between the expansion history and the growth factor of large scale structure   
will be broken. The dark energy equation 
of state $w_{exp}(z)$ which best fits measurements of the expansion 
 will not be consistent with the equation 
of state $w_{growth}(z)$ which best fits measurements the growth. 

The methods and steps we use are as follows.
\\
\textit{i)} We simulate the data for the expansion and the growth 
using a fiducial DGP cosmology (for example, supernova and 
weak lensing data);
\\
\textit{ii)} We use a $\chi^2$ minimization method in order to
find the best fit dark energy model to measurements of the expansion 
(the $\chi^2$ minimization method was discussed in detail in \cite{Upadhye} and was 
shown to give similar results to those from Monte Carlo Markov Chain 
method although the $\chi^2$ minimization can have a better handle on degeneracies \cite{Upadhye});
\\
\textit{iii)} we use the $\chi^2$ minimization in order to
find the best fit dark energy model to measurements of the growth of structure, 
\\
\textit{iv)} Next, we use a standard Fisher matrix approach to calculate 
the confidence regions (or  $\chi^2$ contours) around
the two best fit dark energy models (this standard procedure 
is discussed in detail in references \cite{fisher});
\\
\textit{v)} We look for significant differences between the two effective dark energy
parameter spaces. 
\\
These differences will signal an inconsistency between 
the expansion and the growth factor.
The source of the inconsistency is from point \textit{(i)} where the data
was generated using a DGP model, i.e. from our hypothesis that
the true cosmology is that of a modified gravity DGP model.

The basic idea explored here was also discussed in 
references \cite{lue,Linder2005} but here we propose to implement the 
idea using pairs of cosmological data sets. 
In the next section, we will show how a procedure using
observations from supernovae, weak lensing tomography, 
and the CMB allows one to achieve an observational implementation of  
the consistency test.

Note that our approach is totally different from that of 
\cite{Song2005,Knox2005}.  Their work defined 16 new parameters 
(additional to the cosmological parameters)
describing the distance and growth factor as functions of redshift.
A weak lensing survey and CMB power spectra were then used to
constrain all 16 parameters simultaneously in order to search for
inconsistencies.  
\\
Our work requires no new parameters and 
we explore inconsistencies between constraints obtained from different
pairs of cosmological probes as we illustrate in the next section.
%
%%%%%%%%%%%%%%%%%%%%%%%%%%%%%%%%%%%%%%%%%%%%%%%%%%%%%%%

\section{A procedure to detect signature of modified gravity 
using [SN Ia+CMB] versus [Weak lensing+CMB] observations.} 
We provide here an implementation of the consistency test using 
simulated cosmological observations. This observational test allows one 
to distinguish between dark energy models and modified gravity DGP models.
We assume the true cosmology to be a DGP model 
and we analyze two combinations of simulated data sets using dark 
energy models.
We use the methods and steps indicated in the previous section. 
The procedure we use is as follows. 
\\
\textit{i)} We use a fiducial DGP model (with $\Omega_m=0.27$) and generate supernova magnitudes,
weak lensing convergence power spectrum, and CMB power spectrum.
\\
\textit{ii)} Then, we determine the dark energy model that is the best fit to 
the supernova magnitudes and to the CMB spectrum generated 
using the fiducial DGP model. 
\\
\textit{iii)} Next, we determine the dark energy model that is the best fit to the 
fiducial weak lensing spectrum and to the CMB spectrum. 
\\
\textit{iv)} Then, we compare the allowed regions 
in the $\{\Omega_{de},w_0,w_1\}$ parameter space between the two 
data combinations in order to look for inconsistencies.
\\

{
\begin{figure}[t]
\begin{center}
\includegraphics[width=2.5in,height=3.2in,angle=-90]{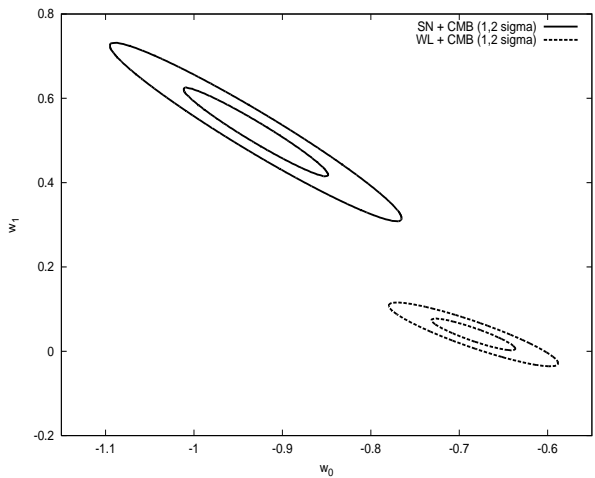}
\caption{\label{fig:eossnwl_a} 
    Equations of state found using two
    different combinations of data sets.  Solid contours are
    for fits to SN Ia and CMB data, while dashed contours are
    for fits to weak lensing and CMB data.  The significant
    difference (inconsistency) between the equations of state found using
    these two combinations is a signature of the DGP
    model and should be detectable by future
    experiments described in sections IV and V.
	The inconsistency is an observational detection
    of the underlying modified gravity DGP model (assumed here to generate the data).
}
\end{center}
\end{figure}
}

The supernova combination contains information on the effect of 
the acceleration on the expansion history, while the weak lensing 
combination contains information on the growth factor in addition 
to the expansion. This translates into a difference between 
the two allowed regions in the dark energy parameter space. 

Our results in Figs. \ref{fig:eossnwl_a}, \ref{fig:eossnwl_b}, and \ref{fig:eossnwl_c}
show that the two allowed regions in the dark energy parameter 
space are significantly different. This signals the expected inconsistency 
between the expansion history and the  
growth rate of large scale structure.

\section{Discussion}

{
\begin{figure}[h]
\begin{center}
\includegraphics[width=2.5in,height=3.2in,angle=-90]{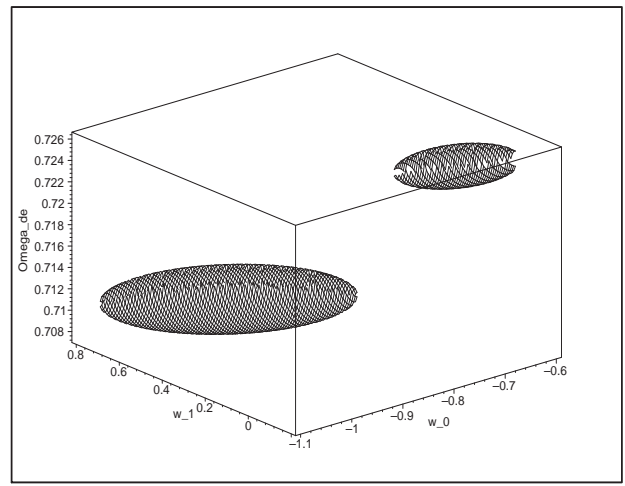}
\caption{\label{fig:eossnwl_b} 
    Dark Energy parameter spaces  ($\Omega_{de}$, $w_0$, $w_1$) found using two
    different combinations of data sets.  The ellipsoid to the left is 
    for fit to SN Ia and CMB data, while ellipsoid to the right is 
    for fits to weak lensing and CMB data.  The significant
    difference (inconsistency) between the parameter spaces found using
    these two combinations is a signature of the DGP
    model and should be detectable by future
    experiments described in sections IV and V.
	The inconsistency is an observational detection
    of the underlying modified gravity DGP model (assumed here to generate the data).
}
\end{center}
\end{figure}
}

The source of the inconsistency in Figs. \ref{fig:eossnwl_a}, \ref{fig:eossnwl_b}, 
and \ref{fig:eossnwl_c} is that we generated the data using a fiducial DGP model but 
then we used dark energy models to fit the data. 
The inconsistency is a consequence of
the fact that the fiducial cosmological model is DGP, 
and therefore it constitutes an observational  
signature of the underlying modified gravity model. 
Fig.\ref{fig:eossnwl_c} shows that the inconsistency  
between the two equations of state persists even when we consider a third term 
in the Taylor expansion of the equation of state (section IIIA). This indicates that 
the test is robust to the functional form used for the
equation of state. 

However, we stress that we did not consider in this study dark energy models 
with couplings or with an unusual sound speed. The study of the effect of the sound 
speed of dark energy on the procedure and also the inclusion of other models of 
dark energy and modified gravity will be considered in subsequent work. 

Also, another point worth exploring in future work is the effect of 
systematic uncertainties. If such a 
procedure is performed using real data and an inconsistency is found, then 
one has to develop methods to test whether the inconsistency is due to physics 
or to systematics in either of the data sets. Further work will
be necessary in order to make this type of test robust and generic.

In summary, the comparison of measurements of expansion history  
and measurements of the growth rate of structure tests the behavior of 
gravity on large scales. 
We proposed a procedure that uses different pairs of cosmological data sets 
in order to explore this comparison and to distinguish 
between some models of dark energy and modified gravity as the
cause of the cosmic 
acceleration. Being able to distinguish between the two possibilities 
is an important step in the quest to understand cosmic acceleration.

{
\begin{figure}[h]
\begin{center}
\includegraphics[width=2.5in,height=3.2in,angle=-90]{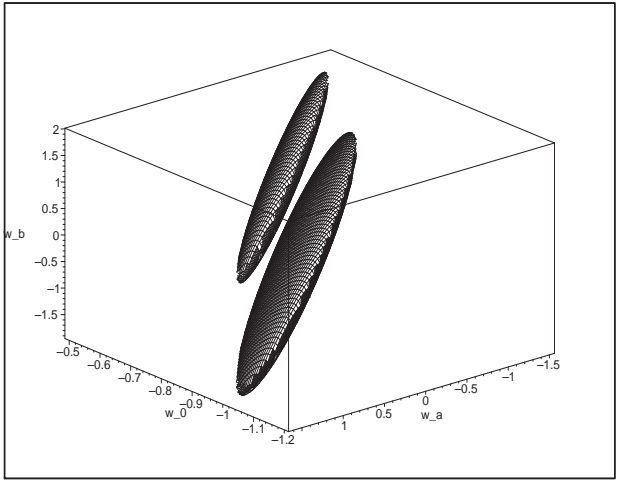}
\caption{\label{fig:eossnwl_c} 
    Three parameter equations of state ($w_0$, $w_a$, $w_b$) (from the Taylor expansion in section IIIA) found using two
    different combinations of data sets.  The ellipsoid to the right is
    for fit to SN Ia and CMB data, while the ellipsoid to the left is
    for fits to weak lensing and CMB data.  The 
    inconsistency between the parameter spaces found using these two combinations is a result of the assumed underlying DGP model and should be detectable by future
    experiments described in sections IV and V.     
}
\end{center}
\end{figure}
}

\acknowledgments
It is a pleasure to thank C. Hirata for useful comments. Part of the analysis
was run on a Beowulf cluster at Princeton University, supported 
in part by NSF grant AST-0216105.
MI is supported by Canadian NSERC and
NASA Award NNG04GK55G. AU is supported by a NSF
Graduate Research Fellowship.
DNS is partially supported by AST-0413793

{}

\begin{thebibliography}{}
%
\bibitem[*]{email}{Electronic Addresses: \\ mishak@utdallas.edu; mishak@princeton.edu}
%
\bibitem{observations1} 
A. G. Riess, {\em{et al.}}, Astron. J. {\textbf{116}}, 1009-1038 (1998);
\bibitem{observations2} 
 S. Perlmutter, {\em{et al.}}, Astrophys. J. {\textbf{517}}, 565-586 (1999);
\bibitem{observations3} 
R. A. Knop, {\em{et al.}}, Astrophys. J. {\textbf{598}}, 102-137
(2003); 
\bibitem{observations4} 
A. G. Riess, {\em{et al.}}, Astrophys. J. {\textbf{607}}, 665-687
(2004);  
\bibitem{observations5} 
C. L. Bennett, {\em{et al.}},
Astrophys. J. Suppl. Ser. {\textbf{148}}, 1 (2003); 
\bibitem{observations6} 
D. N. Spergel, {\em{et al.}},
Astrophys. J. Suppl. Ser. {\textbf{148}}, 175 (2003);
\bibitem{observations7} 
 Seljak et al., Phys.Rev. D{\textbf{71}}, 103515 (2005);
\bibitem{observations8} 
M. Tegmark, {\em{et al.}},  Astrophys. J. {\textbf{606}}, 702-740
(2004).
%
\bibitem{reviews1} 
S. Weinberg, Rev. Mod. Phys., \textbf{61}, 1 (1989);
\bibitem{reviews2} 
S.M.Carroll, W.H. Press and E.L. Turner, \araa, \textbf{30}, 499
(1992); 
\bibitem{reviews3} 
M.S. Turner, \physrep, \textbf{333}, 619 (2000); 
\bibitem{reviews4} 
S.M. Carroll, Living Reviews in Relativity, \textbf{4}, 1 (2001); 
\bibitem{reviews5} 
Varun Sahni, Alexei Starobinsky, Int.J.Mod.Phys., \textbf{D9}, 373
(2000).
\bibitem{reviews6} 
T. Padmanabhan, \physrep, \textbf{380}, 335 (2003). 
%
\bibitem{quintessence1} 
P.J.E Peebles and B.Ratra, Astrophys.J.Lett. \textbf{325} L17 (1988); 
\bibitem{quintessence2} 
B.Ratra and P.J.E. Peebles, Phys.Rev.D \textbf{37} 3406 (1988); 
\bibitem{quintessence3} 
I. Zlatev, L. Wang, and P.J. Steinhardt, Phys. Rev. Lett. \textbf{82} 896 (1999); 
\bibitem{quintessence4} 
P.J.Steinhardt, L. Wang, and I.Zlatev, Phys. Rev. D \textbf{59} 123504 (1999); 
%
\bibitem{DGP} G. Dvali, G. Gabadadze, and M. Porrati, Phys. Lett. B {\textbf{485}} 208-214 (2000).
%
\bibitem{DeffayetDvaliGabadadze2002} C. Deffayet, G. Dvali, G. Gabadadze, Phys.Rev. D {\textbf{65}} 044023 (2002).
%
\bibitem{modifiedgravity1} S. Carroll et al., Phys.Rev. D {\textbf{71}}, 063513 (2005).
%
\bibitem{modifiedgravity2}  S. Nojiri, S. D. Odintsov, Phys. Lett. B {\textbf{631}}, 1 (2005).
%
\bibitem{Upadhye} A. Upadhye, M. Ishak, and P.J. Steinhardt, Phys.Rev. D{\textbf{72}}, 063501(2005). 
%
\bibitem{Ishak2005} M. Ishak, MNRAS, {\textbf{363}}, 469-478 (2005).
%
\bibitem{Chevallier} M. Chevallier, D. Polarski, and A. Starobinsky, Int. J. Mod. Phys. D {\textbf{10}}, 213 (2001).
%
\bibitem{Linder}E. Linder, Phys. Rev. Lett., 90, 091301 (2003).
%
\bibitem{DeffayetLandauRaux} C. Deffayet, {\em{et. al.}}, Phys. Rev. D {\textbf{66}}, 024019 (2002).
%
\bibitem{maor2001}I. Maor, R. Brustein, and P.J. Steinhardt, Phys. Rev. Lett. 86, 6-9 (2001)
%
\bibitem{Deffayet2001} C. Deffayet, Phys. Lett. B {\textbf{502}}, 199-208 (2001).
%
\bibitem{Deffayet2002b} C. Deffayet, Phys. Rev. D {\textbf{66}}, 103504 (2002).
%
\bibitem{lue}  A. Lue, R. Scoccimarro, G. Starkman,.
Phys. Rev. D {\textbf{69}}, 124015 (2004)
%
\bibitem{Koyama} K. Koyama, R. Maartens 
JCAP {\textbf{601}}, 16 (2006)
%
\bibitem{Song2005} Y-S. Song, Phys. Rev. D {\textbf{71}}, 024026 (2005).
%
\bibitem{Ishaketal2004} M. Ishak,  C. M. Hirata, P. McDonald, U. Seljak, Phys. Rev. D {\textbf{69}}, 083514 (2004).
%
\bibitem{Deffayet3}  C. Deffayet, S. J. Landau, J. Raux, M. Zaldarriaga, P. Astier, Phys. Rev. D {\textbf{66}}, 024019 (2002).
%
\bibitem{Sawicki} I. Sawicki, S. M. Carroll, preprint astro-ph/0510364. 
%
\bibitem{LueReport} A. Lue, Phys. Rept. {\textbf{423}},1 (2006). 
%
\bibitem{Fairbairn} M. Fairbairn, A. Goobar, preprint astro-ph/0511029.
%
\bibitem{IshakandHirata2005} M. Ishak and C. Hirata, Phys. Rev. D {\textbf{71}}, 023002 (2005).
%
\bibitem{Ma1999} C.P. Ma, R. R. Caldwell, P. Bode, L. Wang, \apj,  {\textbf{521}}, L1 (1999)
%
\bibitem{linder2003}E. Linder, A. Jenkins, MNRAS, {\textbf{346}}, 573 (2003).
%
\bibitem{1992ApJ...388..272K} N. Kaiser, \apj, {\textbf{388}}, 272 (1992).
%
\bibitem{1997ApJ...484..560J} B. Jain, U. Seljak, \apj, {\textbf{484}}, 560 (1997).
%
\bibitem{1998ApJ...498...26K} N. Kaiser, \apj,  {\textbf{498}}, 26, (1998).
%
\bibitem{Zaldarriaga}M. Zaldarriaga, U. Seljak, cwApJS, 129, 431 (2000). 
%
\bibitem{2000Natur.405..143W}  D. M. Wittman, J. A. Tyson, D. Kirkman, I. Dell'Antonio, G. Bernstein, Nature, {\textbf{405}}, 143 (2000).
%
\bibitem{Albert2005} {J. Albert {\em{et al.}} for the SNAP collaboration}, astro-ph/0507460 (2005).
%
\bibitem{smith2003} {R.E.Smith, {\em{et al.}}}, MNRAS, {\textbf{341}}, 1311 (2003).
%
\bibitem{Hu1} W. Hu, Astrophys.J. 522, L21-L24 (1999).
%
\bibitem{Hu2} W. Hu,  Phys.Rev. D{\textbf{66}}, 083515 (2002).
%
\bibitem{fisher} Press et al., Numerical Recipes in C, Cambridge University Press (1992); S. Dodelson, Modern Cosmology, Academic Press (2003); Tegmark, Taylor, and Heavens, Astrophys.J. 480, 22 (1997). 
%
\bibitem{Linder2005} E. Linder,  Phys. Rev. D{\textbf{72}}, 043529 (2005).
%
\bibitem{Knox2005} L. Knox, Y.-S. Song, J.A. Tyson, astro-ph/0503644.
%
\end{thebibliography}
\end{document}